# Defect-induced oxygen adsorption on graphene films

Tianbai Li and Jory A. Yarmoff*

*Department of Physics and Astronomy, University of California, Riverside, Riverside CA 92521*

## Abstract

Although defects on graphene can degrade electron transport and its ability for use as a protection layer, they can also be helpful to tailor the local properties or activate new sites for particular adsorbates. Here, carbon vacancy defects are formed in graphene films on Ru(0001) using low energy $Ar^+$ bombardment and the materials are then reacted at room temperature with oxygen ($O_2$). Helium low energy ion scattering shows that no oxygen attaches to the intact graphene layer. When isolated single carbon vacancy defects are present, oxygen adsorbs molecularly at the defect sites and intercalates beneath the graphene overlayer after post-annealing at 600 K. When the defects are large enough to consist of open areas of bare substrate, the oxygen dissociatively chemisorbs to the Ru. This work shows that the adsorption depends on the size of the surface vacancies, and that it is important to have defect-free graphene when using it as a protection layer.

*Corresponding author, E-mail: yarmoff@ucr.edu, phone: +1 (951) 827-5336

## 1. Introduction

The epitaxial growth of uniform and large areas of graphene (Gr) with excellent quality is now routinely achieved by chemical vapor deposition (CVD) on transition metal surfaces [1-5]. The inertness and high thermal stability of graphene make it a good candidate for use as a protective layer, especially for transition metals [6,7]. Nevertheless, in the transferring of graphene or in the fabrication of devices, it is inevitable that some defects, most likely carbon vacancies, will be introduced. It can be expected that those defects will leave unoccupied sites at which contaminants can absorb and thus degrade the graphene's performance over a large scale. Such a deviation from a perfect graphene film can, however, be useful in some applications. For example, such defects can be used to tailor the local properties of graphene and achieve new functionalities [8]. Although defects in graphene have been well studied and widely used, the relationship between the defects and the adsorbates on the surface is still not clear.

One of the most important adsorbates is oxygen due to its high reactivity and abundance in the atmosphere. In general, at room temperature $O_2$ molecules adsorb dissociatively on metal surfaces forming chemical bonds to surface atoms [9-11]. Oxygen can adsorb molecularly via van der Waals forces at sufficiently low temperatures [11-14]. Molecular adsorption, however, can be observed at a relatively high temperature in some rare cases. For instance, a small amount of oxygen can adsorb in molecular form on certain Si surfaces at room temperature in which it serves as precursor for dissociative chemisorption during the process of forming fully oxidized $SiO_2$ [15] or as a minority species after the initial formation of a surface oxide [16].

There are, however, many reports of the adsorption of molecular $O_2$ in the presence of defects. For example, $O_2$ has been shown experimentally to molecularly chemisorb at two types of adsorption sites on $TiO_2$ with oxygen vacancies at temperatures of 150 K and 230 K [17]. Also, when defects are created on Ag(111) by exposure to a high dose of chlorine, molecular oxygen can stick to those defects at room temperature [18]. In addition, DFT calculations have shown that defects on $MoS_2$ and doped boron nitride surfaces, which are two-dimensional materials that have similar structures as Gr, are active sites that enable the uptake of $O_2$ and $Cl_2$ molecules with a much lower adsorption energy [19,20]. It is thus important to explore the possible adsorption of molecules on defected graphene at temperatures that are higher than those usually associated with physisorption on surfaces.

In this paper, the relationship between defects on graphene and adsorbed $O_2$ is studied with $He^+$ low energy ion scattering (LEIS) [21]. As a highly surface sensitive tool, LEIS has been previously applied to measure the impurities in graphene overlayers [22] and intercalation of molecules underneath the Gr overlayer [23]. In particular, by adjusting the scattering angle, LEIS spectra collected from Gr films can detect only the outermost atoms or can also detect intercalated species and uncovered substrate atoms [23]. In our previous study [23], a comparison of spectra collected at different scattering angles was used to show that $O_2$ molecules intercalate between the Gr overlayer and the Ru(0001) substrate when exposed to $O_2$ at 650 K, and that no oxygen adsorbs on the surface. Furthermore, by comparing the desorption temperature of the intercalated oxygen to that of oxygen chemisorbed directly on Ru(0001), it was concluded that the intercalated oxygen remains in molecular form.

This work presented here details the role of carbon vacancies in promoting adsorption on Gr. As reported in Refs. [24,25], $Ar^+$ sputtering can create single carbon vacancies on Gr/Pt(111) and graphite, as confirmed by STM images. Here, defects are introduced on Gr/Ru(0001) by 50 eV $Ar^+$ pre-sputtering before the material is exposed at room temperature to $O_2$. Although a complete graphene layer is inert to adsorption at room temperature, it is found that $O_2$ does adsorb on pre-sputtered Gr/Ru(0001). In addition, it is observed that the size of the defects affects the form of the adatoms and their stability. For a light sputtering that generates isolated single C vacancy defects, oxygen adsorbs molecularly on the defect sites and diffuses to become intercalated between Gr and the Ru(0001) substrate following post-annealing at 600 K. If the vacancies are large enough to produce open areas of exposed substrate, however, then $O_2$ dissociates to form strong O-Ru metal bonds.

## 2. Experimental procedure

The experiments are performed in an ultra-high vacuum (UHV) chamber with a base pressure of $4\times10^{-10}$ Torr. An $Ar^+$ ion sputter gun (Varian) is used to clean the Ru substrate and introduce defects into the Gr overlayer. Sapphire leak valves are used to introduce Ar gas for sputtering and other gasses for the growth of the Gr overlayer and the introduction of oxygen molecules. For the analysis of the surface and the confirmation of the cleanliness, the chamber is equipped with low energy electron diffraction (LEED) optics (Varian) and the LEIS setup, which is described below. The sample is mounted on a holder attached to the foot of a manipulator that enables x-y-z motion and rotation about both the polar and azimuthal angles. An e-beam heater filament is mounted behind the sample holder that can be floated at a negative high voltage to heat the sample up to 1400 K. The temperature of the sample is monitored with K-type thermocouples that are spot-welded in the vicinity of the sample. There is a Faraday cup with a 1 mm diameter entrance hole attached at the end of the manipulator foot for accurately measuring the ion beam sizes and fluxes.

The cleaning of the ~1 cm diameter Ru(0001) sample is performed using a standard IBA and chemical treatment, as reported in the literature [26,27]. A 30 min 500 eV $Ar^+$ ion sputtering is first applied to the Ru(0001) sample at a flux of $4\times10^{13}$ ions $sec^{-1}$ $cm^{-2}$ with a beam size of $3\times3$

cm$^2$. The sample is then annealed under 4×10$^{-8}$ Torr of O$_2$ at 1100 K for 8 min to remove adsorbed carbon, followed by a flash annealing at 1300 K for 2 min under UHV to remove the remaining carbon-containing contaminants and residual oxygen. This IBA/chemical treatment is normally repeated several times to acquire a clean and well-ordered surface. The cleanliness and the periodicity of the surface are confirmed with LEIS and LEED, respectively.

The graphene overlayer is grown through a chemical vapor deposition (CVD) method [27]. The cleaned Ru surface is heated to 900 K and then exposed to 1.5×10$^{-7}$ Torr of ethylene for 5 min, followed by annealing under vacuum at 1200 K for 1 min and then slowly cooling down to 450 K for another 5 min. This process is repeated until the surface is fully covered with a continuous monolayer of graphene, which typically requires about 4 cycles. The quality of the Gr/Ru(0001) overlayer is monitored with LEIS and LEED.

Defects are formed via Ar$^+$ sputtering of Gr/Ru(0001). To gently remove carbon from the graphene, a low beam energy (around 50 eV) is employed and the beam is defocused to produce an average flux of 6.7×10$^9$ ions sec$^{-1}$ cm$^{-2}$. To study the effects of defect size, two different sputtering times are used. For the "light" sputtering, the beam is applied for 3 min so that the total fluence is 1.2×10$^{12}$ ions cm$^{-2}$. Considering that the lattice constant of free-standing graphene at room temperature is about 2.45 Å [28], it is estimated that 1 out of 167 Gr carbon atoms, or 0.6% of the surface carbon, is impacted during this light sputtering. The second defect formation involves a 1-hour sputtering, which corresponds to a fluence of 2.4×10$^{13}$ ions cm$^{-2}$, so that 1 out of every 8 Gr carbon atoms is impacted.

After the introduction of defects in Gr/Ru(0001), O$_2$ exposures are performed with the sample held at room temperature. Exposures are given in units of Langmuirs (1 L = 1x10$^{-6}$ torr sec). Additional post-annealing under UHV is also performed after the exposures.

The helium low energy ion scattering is performed with a differentially pumped ion gun (PHI model 04-303) that produces a beam diameter of 1.6 mm with a total sample current of 1.5 nA. The scattered ions are collected by a Comstock AC-901 hemispherical electrostatic analyzer (ESA) mounted on a rotatable platform in the UHV chamber, which allows the scattering angle to be adjusted. The ESA collects only those projectiles that remain ionic after scattering. A specular geometry is used for all the spectra collected in this paper in which the incident and outgoing angles are always equal with respect to the surface normal. No detectable damage of the Gr overlayer due to the $He^+$ ion beam occurs within the time it takes to collect 5 successive spectra [23]. Thus, to absolutely avoid effects of any beam damage during LEIS, the sample is re-prepared after the collection of every 3 spectra.

## 3. Results

The primary tool used for these experiments is LEIS [21]. The scattering process can be analyzed classically because the de Broglie wavelength of low energy ions (1-10 keV) is small. In addition, due to the small ratio between the scattering cross sections and the interatomic spacings, low energy ion scattering from a solid can be analyzed with the binary collision approximation (BCA) in which it is assumed that the projectile interacts with only one surface atom at a time [29]. The most significant features in LEIS spectra are the single scattering peaks (SSP) that represent projectiles that experience one hard collision with a single target atom before scattering from the surface [21]. The position of a SSP depends on the mass of the target atom and the scattering angle, while the area is proportional to the number of target atoms that are directly visible to the incoming ions and the detector. In this experiment, the areas of the SSPs are computed by integrating the peaks after subtracting the background of multiply scattered projectiles, which is modeled as a polynomial by fitting the shape of the region surrounding the SSPs.

In the present measurements, a very light projectile, such as helium, is needed to enable backscattering from C and O surface species. Helium LEIS also has an extremely high surface sensitivity due to Auger neutralization (AN), which is an irreversible process that dominates for noble gas projectiles [21,30]. In the process of AN, most of the projectiles that collide with deeper layer atoms remain neutralized and are not detected by the ESA. Thus, the spectra consist primarily of single scattering events from the outermost few atomic layers. There is also, however, a strong matrix effect for helium ions scattered from graphitic carbon making it difficult to detect scattered $He^+$, as reported in the literature [22,31,32]. The reason is that He ions undergo a quasi-resonant neutralization in conjunction with AN when scattering from graphitic carbon that leads to a very high neutralization probability [33]. This effect is particularly significant for primary beam energies below 2500 eV. To avoid the matrix effect and provide a detectable signal, a 3000 eV $He^+$ beam energy is employed here.

Figure 1 shows $He^+$ LEIS spectra collected from Gr/Ru(0001) at a 45° scattering angle after various treatments. At this small scattering angle, the incident and scattered ions are 22.5° from the surface plane so that atoms positioned below the graphene overlayer are completely shadowed by the overlayer, which leads to a signal that only probes the Gr surface and any adsorbates attached to it [23].

Spectrum (a) was collected from the clean, as-prepared Gr/Ru(0001) and shows only a single SSP at 2450 eV, which represents carbon in the Gr overlayer. In addition, the LEED pattern collected from this surface displays a Moiré pattern, as reported previously [23,34], which indicates that the overlayer is a single crystal that forms a superlattice with the substrate due to their different lattice parameters. These data indicate that a complete Gr film covers the surface without a measurable number of defects.

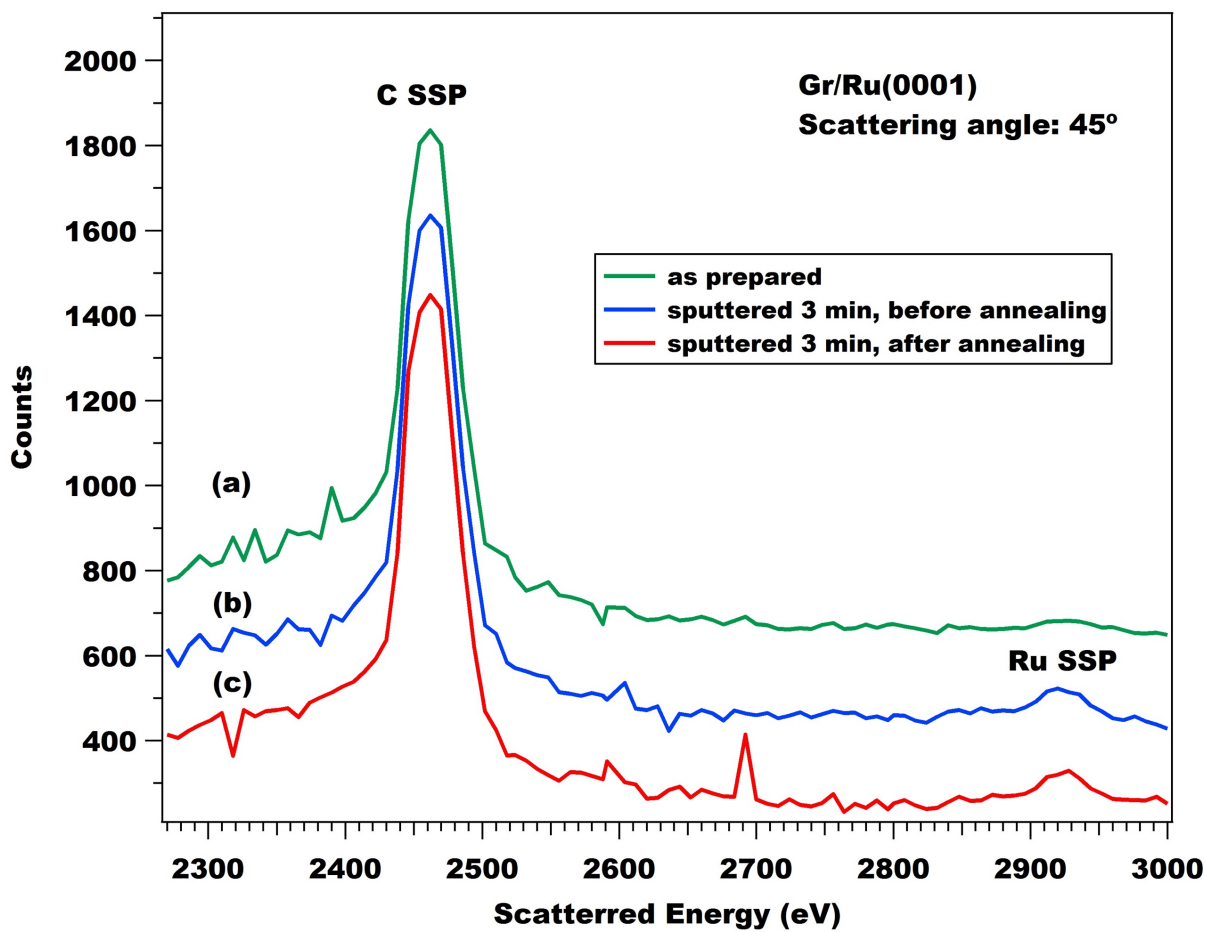

**Figure 1**. 3000 eV He$^+$ LEIS spectra collected at a 45° scattering angle from (a) as prepared Gr/Ru(0001), (b) Gr/Ru(0001) pre-sputtered for 3 min (fluence = $1.2\times10^{12}$ ions/cm$^2$), and (c) after an additional annealing to 1000 K for 5 min. The relevant SSP's are indicated and the y-axes are offset from each other for clarity.

Spectrum (b) in Fig. 1 was collected after the sample was sputtered by 50 eV Ar$^+$ for 3 min, which is the "light" sputtering that corresponds to a fluence of $1.2\times10^{12}$ ions cm$^{-2}$. The spectrum displays a small Ru SSP at around 2920 eV, as well as the C SSP, which indicates a small number of carbon vacancies that reveal the Ru substrate. The LEED pattern observed after the 3 min sputtering does not change, signifying that the small number of defects created is insufficient to affect the alignment between the Gr overlayer and the Ru substrate. The cross section for 3000 eV He$^+$ scattering from Ru is about a factor of 16 larger than for scattering from C [35], so that the relative changes of the Ru SSP area are much larger than those of the C SSP and the number of revealed Ru atoms is rather small. If it is assumed that each defect reveals a single Ru substrate atom and that the neutralization probability is the same for scattering from C and Ru, the ratio of the Ru SSP to the C SSP, after correction for the cross section difference, implies that approximately only 0.4% of the C atoms are removed by the "light" sputtering. This number is similar to the estimate that 0.6% of the C atoms were impacted by Ar$^+$, and thus suggests that approximately one defect is formed by each impact. The small number of missing C atoms is consistent with the notion that the light sputtering creates isolated, single vacancy defects that are randomly positioned across the surface [24,25].

The sputtered surface was then annealed at 1000 K for 5 min, and spectrum (c) in Fig. 1 was collected. It is seen that the areas of the C and Ru SSPs do not change with annealing, which infers that the vacancies induced by the light sputtering of Gr/Ru(0001) do not recover on their own. Thus, any possible restoration of the graphene lattice caused by post-annealing can be neglected.

After exposure to O$_2$, 3000 eV He$^+$ LEIS spectra were collected at a 45° scattering angle to probe the outermost Gr layer and any possible adsorbates, as shown in Fig. 2. Spectrum (a) was collected from fully-covered Gr/Ru(0001) after an 8000 L oxygen exposure with the sample at room temperature. The only visible peak in the spectrum is the C SSP, which indicates that oxygen does not adsorb on top of the carbon atoms in a continuous graphene layer. Spectrum (b) was collected from Gr/Ru(0001) after a 3 min light sputtering followed by a 1500 L O$_2$ exposure at room temperature. A new peak appears at 2580 eV that represents the O SSP, while the area of C SSP does not change. A small Ru SSP also appears, but it is smaller than the Ru SSP from the sputtered surface prior to O$_2$ exposure shown in Fig. 1(b). It is thus inferred that randomly distributed single vacancies on the graphene surface enable the adsorption of oxygen, even at room temperature, and that the adsorbed oxygen initially sits on or near the defects so that some of the underlying Ru is shadowed. The sample was then annealed to 600 K for 10 minutes and re-measured to produce spectrum (c) of Fig. 2. It is found that annealing causes the oxygen peak to disappear and the Ru signal to increase. This indicates that the adsorbed oxygen either desorbs from the surface or diffuses underneath the Gr after heating so that the Ru sites near the carbon vacancy defects are less shadowed.

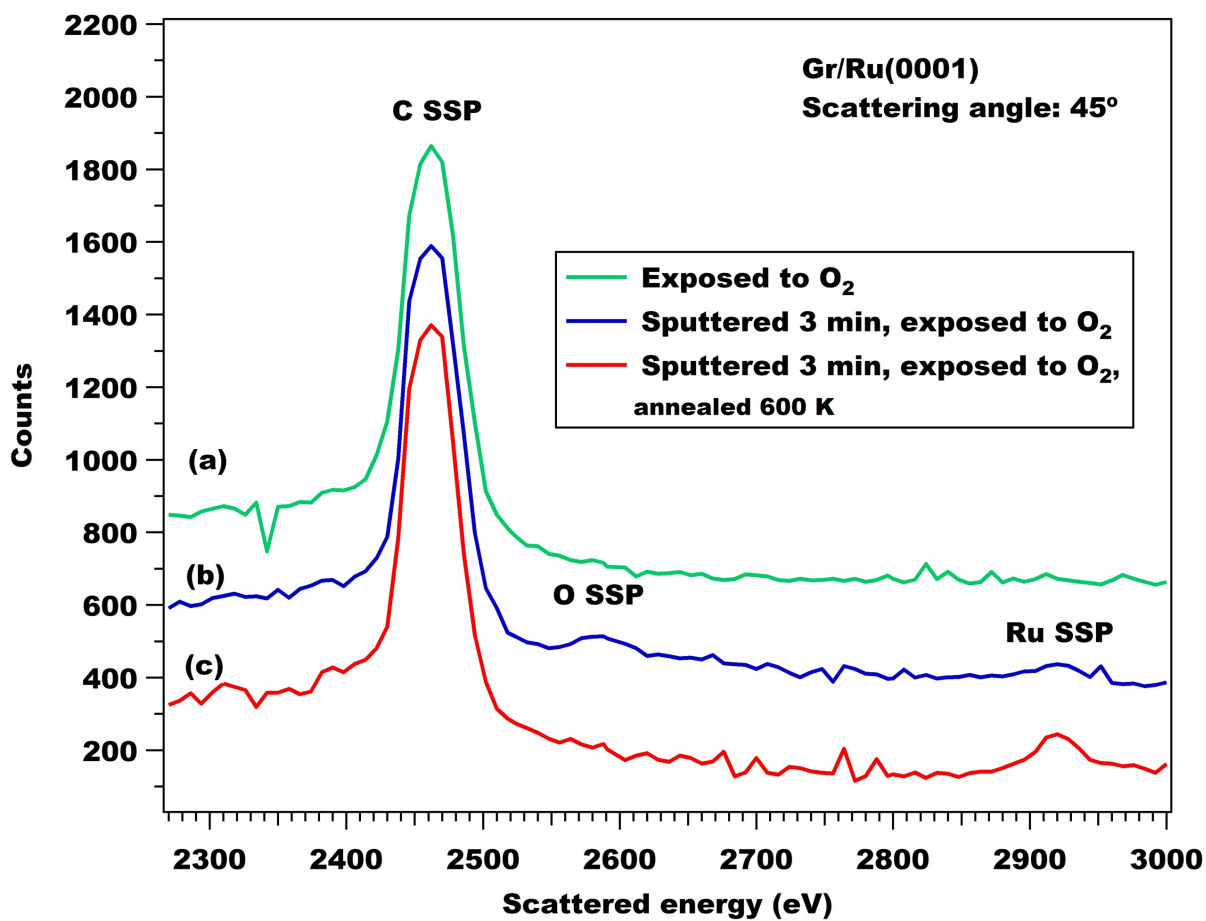

**Figure 2**. 3000 eV He$^+$ LEIS spectra collected at a scattering angle of 45° from (a) as-prepared Gr/Ru(0001) exposed to 8000 L of O$_2$ at 300 K, (b) Gr/Ru(0001) pre-sputtered for 3 min and then exposed to 1500 L of O$_2$ at 300 K, and (c) after additional annealing at 600 K for 10 min. The relevant SSP's are indicated and the spectra are offset from each other for clarity.

To further study oxygen adsorption on Gr/Ru(0001) with isolated C vacancies, the above LEIS measurement was repeated at a 120° scattering angle, as shown in Fig. 3. A larger scattering angle enables some ions to pass through the graphene layer making it possible to detect both the surface atoms and those that have intercalated beneath the Gr overlayer. Note that the SSPs occur at lower energies than in Figs. 1 and 2 because of the larger scattering angle. From spectrum (a) in Fig. 3, it is found that no $O_2$ intercalates underneath a pristine Gr film at room temperature, in contrast to what it does at elevated temperature [23,36]. Note that the small Ru SSP is caused by the $He^+$ ions that can penetrate the overlayer and backscatter at this angle. Similar to spectrum (b) in Fig. 2, spectrum (b) in Fig. 3 shows both O and Ru SSPs after a 1500 L room temperature oxygen exposure. After the additional annealing to 600 K, however, the oxygen peak in Fig. 3(c) decreases but does not disappear as it did in Fig. 2. This indicates that the oxygen has diffused from the previous adsorption site at a carbon vacancy to a new position that is between the Gr overlayer and Ru substrate, i.e., most of the oxygen intercalates when annealed rather than desorbs. The small decrease in the O SSP area is likely due to the neutralization probability being larger for $He^+$ scattered from buried oxygen.

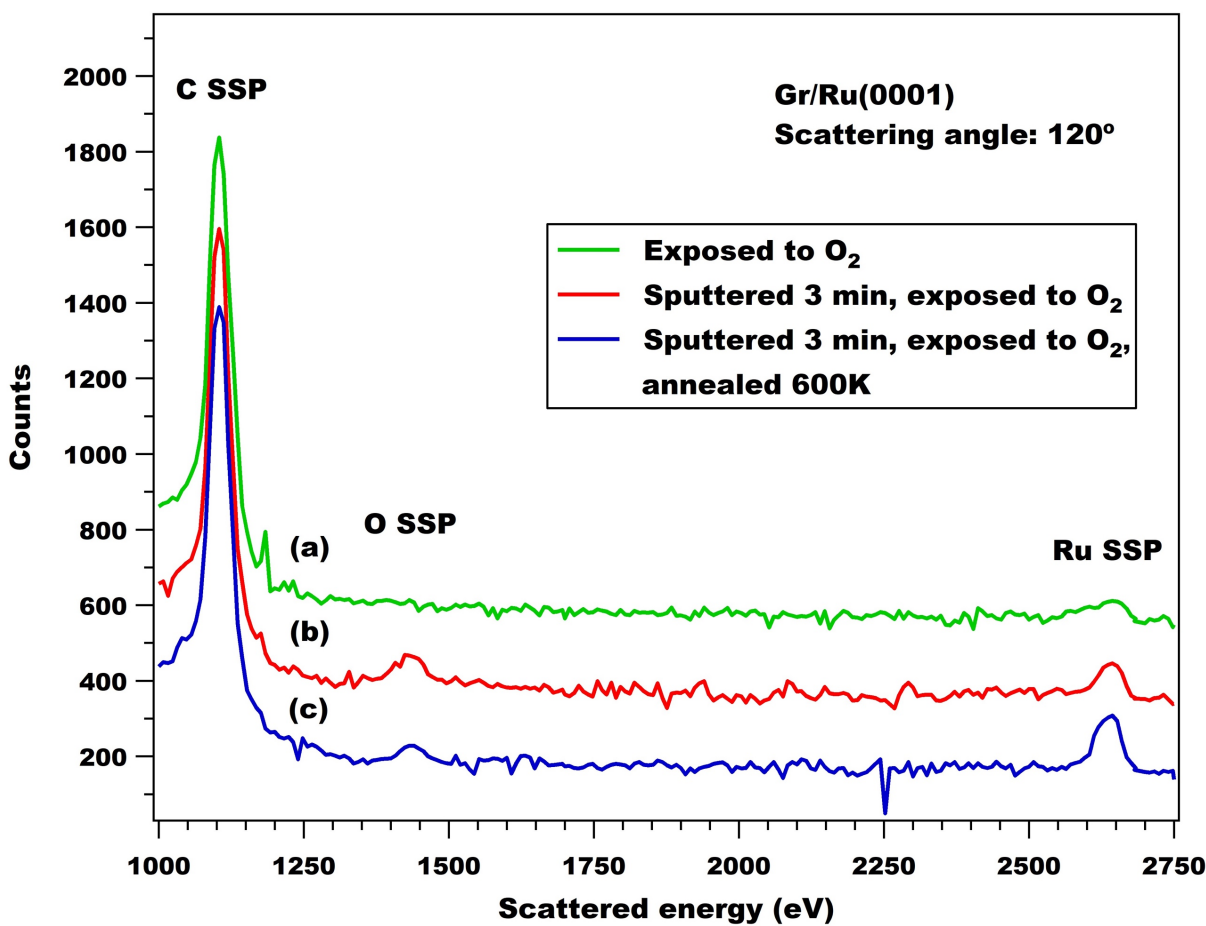

**Figure 3**. 3000 eV He$^+$ LEIS spectra collected at a scattering angle of 120° from (a) as-prepared Gr/Ru(0001) exposed to 8000 L of O$_2$ at 300 K, (b) Gr/Ru(0001) pre-sputtered for 3 min and then exposed to 1500 L of O$_2$ at 300 K, and (c) after an additional annealing at 600 K for 10 min. The relevant SSP's are indicated and the spectra are offset from each other for clarity.

To explore the relationship between the oxygen adsorption and defect size, further measurements are performed in which larger defects are produced by using a longer sputtering time. Figure 4 shows spectra collected from a Gr/Ru(0001) sample that was sputtered for 1 hour, and then exposed to $O_2$ at room temperature and subsequently annealed. The LEIS spectra are collected at a 45° scattering angle so that any intercalated oxygen buried beneath the graphene layer is not visible. Spectrum (a) was collected from the Gr/Ru(0001) sample following the 1 hour sputtering. The Ru SSP is much larger than that in Fig. 1(b) indicating that more Ru target atoms are visible to the incoming ions due to the larger vacancy defects. An analysis of the C/Ru SSP ratio suggests that approximately 2.3% of the carbon is removed by the 1 hour sputter. In addition, the Moiré pattern of Gr/Ru(0001) becomes blurry after the 1 hour sputtering indicating that the supperlattice structure is partially destroyed. The LEIS data collected after the sputtered sample is exposed to 1500 L of $O_2$ at room temperature are shown in spectrum (b). The presence of the oxygen peak, the unchanged intensity of the carbon peak and the decrease of the size of the Ru peak suggest that oxygen adsorbs in the large defects and partially shadows the Ru underneath. As seen in spectrum (c), after annealing the sample to 600 K, both the O and the Ru SSPs do not change intensity, which is in contrast to the behavior after the 3 min sputtering seen in Fig. 2. Considering that oxygen chemisorbed on bare Ru(0001) is very stable and does not desorb until being heated to 1200 K [23,36], it can thus be inferred that most of the $O_2$ dissociatively chemisorbs as atomic oxygen to bare areas of the Ru substrate produced by the prolonged sputtering, and does not react with the Gr film.

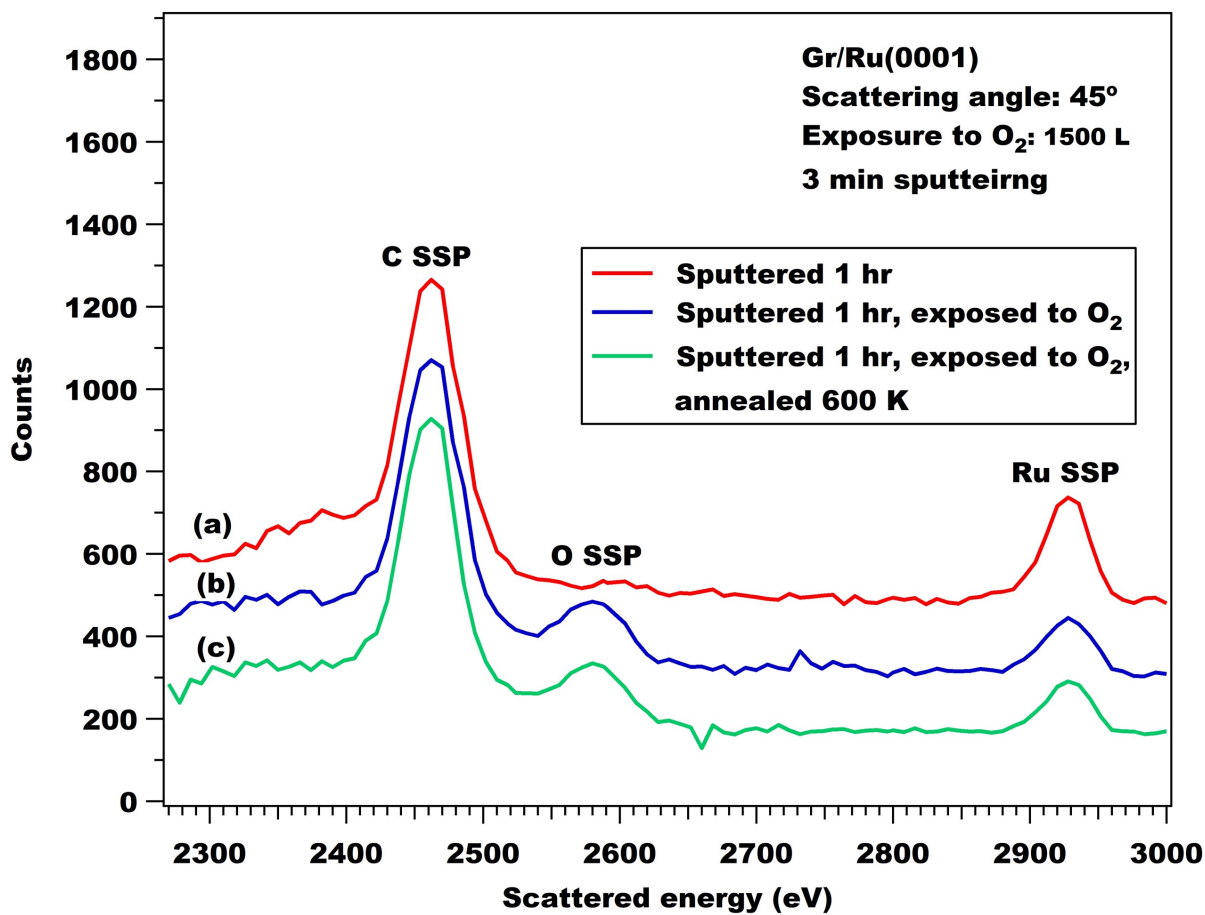

**Figure 4**. 3000 eV He[+] LEIS spectra collected at a scattering angle of 45° from (a) Gr/Ru(0001) sample after a 1 hour sputtering, (b) the sputtered sample after exposure to 1500 L of $O_2$ at 300 K, and (c) after an additional annealing at 600 K for 10 min. The relevant SSP's are indicated and the spectra are offset from each other for clarity.

**4. Discussion**

An implicit assumption in this analysis is the intrinsic stability of the single C vacancy defects on Gr/Ru(0001), which is based on the data shown in Fig. 1 after annealing. Indeed, free-standing graphene is able to reconstruct its lattice in the presence of defects under certain conditions [8]. As a good example, the Stone-Wales intrinsic defect (4 hexagons are transformed into 2 pentagons and 2 heptagons) can occur without any removed or added C atoms after heating graphene over 1000 K or through mechanical strain [8,37]. For a single vacancy in graphene, the formation of this reconstruction is not easy, however, as the energy required has a calculated value of 7.5 eV [8,38], which is much higher than the vacancy formation energy of many other materials [39,40]. Once a single vacancy is formed by rapid quenching from high temperature, collision or irradiation, one carbon atom is permanently lost which can lead to the formation of 5-membered and 9-membered rings [25,41]. That structure is very stable so that the restoration of the single vacancy does not occur, as is suggested by the spectra of Fig. 1. Nevertheless, migration of single vacancies is possible when heated. As reported in Ref. [38], the calculated migration energy is about 1.3 eV, which should allow migration at temperatures slightly above 200°C. This migration of the defects cannot be detected in the LEIS spectra of Fig. 1 after annealing, however, because of the much larger scale of the beam size (1.6 mm) compared to the graphene lattice, and because the diffusion of the single vacancies is completely random.

When a large number of C atoms are removed from the surface by sputtering, a reconstruction could occur that involves bending and warping of the Gr layer due to its reduced surface area and the formation of open regions that have unsaturated bonds around their circumference, as occurs for free standing Gr [8]. For Gr films on a substrate treated with a lengthy Ar$^+$ sputtering, it is likely that the carbon vacancies migrate and coalesce after many C atoms are removed, even at room temperature, due to the instability of the film when it is missing too many carbon atoms, which leads to the formation of continuous areas that are void of Gr within an otherwise crystalline Gr overlayer. Such a case was observed in the aberration-corrected TEM images of monolayer graphene after a prolonged irradiation to a high energy electron beam [42], where 11% of the carbon was removed and continuous defects formed. The estimate above from the LEIS data is that about 2.3% of the surface consists of large areas that are void of carbon in which the bare Ru(0001) substrate is revealed.

It is thus shown in this paper that the size of the vacancies on Gr/Ru(0001) makes a difference in the adsorption and subsequent diffusion of the oxygen. To systematically compare how the defect size affects the adsorption of oxygen, the Ru SSP areas collected at 45° from the clean surface and following the sputtering, oxygen adsorption and annealing steps are plotted in Fig. 5. The triangles (a) correspond to the 1 hour pre-sputtering, and the circles (b) correspond to the 3 min pre-sputtering. It is observed that the areas of both of the Ru SSPs increase after the pre-sputtering due to the removal of C atoms, but the increase is larger following the longer sputtering time, as expected. Following a 1500 L oxygen exposure at room temperature, both Ru SSP areas decrease, but the magnitude of the decrease is different for the two sputtering times. This could be attributed to the number of available adsorption sites being different for the two cases or that the efficiency of oxygen adsorption at the different sites is not the same. After an additional annealing at 600 K for 10 min, little change of the Ru peak area is observed in (a), but the Ru SSP area in curve (b) rises close to what it was right after the pre-sputtering, suggesting that the isolated vacancy sites are void of oxygen following annealing.

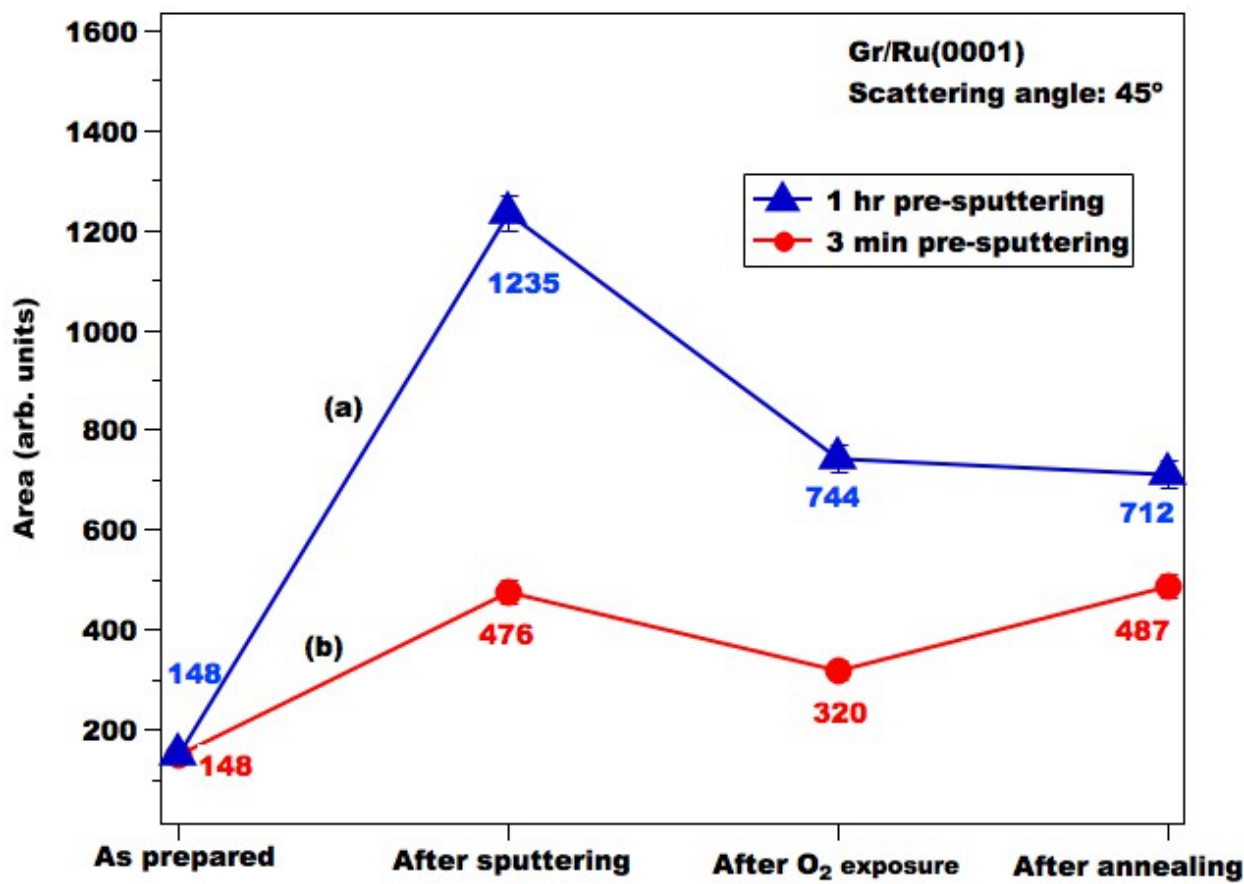

**Figure 5**. The Ru SSP areas of (a) 3 min pre-sputtered Gr/Ru(0001) and (b) 1 hour pre-sputtered Gr/Ru(0001) following the indicated treatments. The $O_2$ exposures are 1500 L, the annealing temperature is 600 K, and the 3000 eV $He^+$ LEIS spectra are collected at a 45° scattering angle.

The likely explanation for the behavior of lightly sputtered Gr/Ru(0001) is that the oxygen remains molecular when attached at a single carbon vacancy site. The adsorption of $O_2$ in molecular form on surface defects has been demonstrated previously [17,19,20]. For example, it has been shown from STM images that molecular oxygen species adsorbs at two different defect sites on rutile $TiO_2$ at a temperature as high as 230 K [17]. Such a temperature is much higher than it is needed to maintain molecular adsorption on the perfect TiO$_2$ surface (60 K). In Ref. [19], two stable configurations of O$_2$ adsorption on MoS$_2$-V$_s$ (with one sulfur vacancy) and MoS$_2$-V$_{s2}$ (with two sulfur vacancies) were proposed using first principles DFT calculations. The corresponding adsorption energies are -1.822 eV and -1.687 eV, respectively, thus producing a more stable structure than the perfect MoS$_2$ surface. Additionally, a similar molecular adsorption of O$_2$ was proposed for doped boron nitride with adsorption energies ranging from -1.13 eV to -5.677 eV, depending on the type of dopant [20]. The adsorption energy for molecular oxygen on pristine boron nitride is only -0.008 eV, however, making it difficult to occur except at extremely low temperatures [20]. Thus, it can be inferred that single C vacancy defects in Gr activate a new adsorption site that stabilizes O$_2$ molecules.

A question to be addressed is whether these conclusions concerning the adsorption of species at defect sites for Gr films can be extended to free graphene. For example, the adsorption of molecular oxygen at single C vacancy sites could involve only bonding to other carbon atoms while the Ru substrate has no effect. The adsorption energy of molecular oxygen on free-standing graphene is, however, much affected by the type of defect per the DFT calculations of Refs. [43,44]. The energy of molecular oxygen adsorption on a Stone–Thrower–Wales defect is -0.17 eV, which should only be observed at very low temperatures [43]. Molecular adsorption at a divacancy site of free-standing graphene, however, has an adsorption energy about -8.48 eV, which should survive even above room temperature [44]. Thus, in the present case, the Ru(0001) substrate might affect the stability of the adsorbates at single vacancy defects. In addition, the difference of lattice sizes between the substrate and deposited graphene leads to a corrugation of the overlayer [45], so that those carbon vacancies residing at different sites with respect to the substrate might not be equivalent. This suggests that information about adsorption at single C vacancy sites on Gr films may not transfer to free-standing Gr, and more work is needed to determine whether or not this is the case.

In the present experiment, the intercalated oxygen that diffuses at 600 K from single carbon vacancy adsorption sites on Gr/Ru(0001) to become intercalated completely desorbs when the sample is heated to 1000 K for 10 min. This indicates that the intercalated oxygen is molecular as, according to the analysis in Ref. [23], atomic oxygen chemisorbed on Ru doesn't desorb until 1200 K. It can therefore be inferred that the $O_2$ molecules are initially physically adsorbed at the vacancy sites before the post annealing and that such a physical bond is so weak that when the sample is gently heated to 600 K, the physisorbed $O_2$ molecules migrate to find more stable sites as intercalates. Certainly, some oxygen could also desorb from the surface during the annealing, but the spectra in Fig. 3 do not show a significant change in the size of the O SSP after annealing.

A one hour pre-sputtering of Gr/Ru(0001) leaves larger open areas of the substrate. The oxygen adsorbed on these exposed substrate areas is more stable and does not diffuse even after being heated to 600 K. As is well known, the dissociative adsorption of $O_2$ occurs readily on transition metals at room temperature [10,11]. For Ru(0001), this dissociative adsorption only requires an energy of about 0.15-0.19 eV, which is obtained at room temperature [46]. On the graphene overlayer, however, the dissociation energy of $O_2$ is 2.71 eV [47], making graphene a good protective layer for transition metals at temperatures below 700 K [6,7]. Thus, it is likely that $O_2$ easily dissociates to form strong O-Ru bonds in the open areas of bare Ru(0001) and does not attach to directly to, or intercalate beneath, the Gr.

## 5. Conclusions

Gr films on transition metal substrates are an important class of carbon-based materials. Uniform and completely covered graphene layers can be used to prevent $O_2$ from reacting with transition metals. It is shown here that $O_2$ does not adsorb on the top of Gr/Ru(0001) nor does it intercalate between the Gr and the substrate at sample temperatures below 450 K [36]. Oxygen can, however, adsorb at Gr carbon vacancy defects even at room temperature. For isolated C vacancies, the oxygen adsorbs molecularly. If the defects are large enough, then oxygen dissociates and chemically bonds directly to the revealed areas of bare metal substrate. This indicates that defects activate new adsorption sites on Gr/Ru(0001) that are not present on pristine graphene and that the adsorption sites of the oxygen and the stability of the adsorbates are very much affected by the size of the defects. These results demonstrate that it is necessary to maintain a complete graphene overlayer when it is being used as a protection layer, or in devices for which the Gr is part of a heterostructure.

## 6. Acknowledgements

This material is based upon work supported in part by the National Science Foundation under CHE - 1611563.